\documentclass[useAMS,usenatbib]{mn2e}
\usepackage{times}
\usepackage{amssymb}
\usepackage{epsfig}

\newcommand{\rmd}{\mathrm{d}}

\begin{document}

\title{On de-Sitter Geometry in Crater Statistics}
\author[G.W. Gibbons and M. C. Werner] {G.W.~Gibbons$^1$ and
M. C.~Werner$^2$\thanks{E-mail:marcus.werner@ipmu.jp}\\ 
$^1$Department of Applied Mathematics and Theoretical Physics, University of Cambridge,
Wilberforce Road, Cambridge, CB3 0WA, United Kingdom\\
$^2$Kavli Institute for the Physics and Mathematics of the Universe, University of
Tokyo, 5-1-5 Kashiwanoha, Kashiwa, 277-8583, Japan, \\
$^{\phantom{}}$ Department of Mathematics, Duke University, Durham, NC 27708, USA}
\date{This draft \today}
\pagerange{\pageref{firstpage}--\pageref{lastpage}} \pubyear{0000}
\maketitle
\label{firstpage}

\begin{abstract}
The cumulative size-frequency distributions of impact craters on planetary bodies in the solar system appear to approximate a universal inverse square power-law for small crater radii. In this article, we show how this distribution can be understood easily in terms of geometrical statistics, using a 
de-Sitter geometry of the configuration space of circles on the Euclidean plane and on the unit sphere. The effect of crater overlap is also considered.
\end{abstract}
\begin{keywords}
planets and satellites: general--cosmology:miscellaneous--methods:statistical
\end{keywords}

\section{Introduction}
When observations by the spacecraft Dawn enabled the first detailed survey of impact craters on the astroid 4 Vesta recently, it was noted that the cumulative size-frequency distribution can be approximated by a power-law with (double logarithmic) slope $-1.9\pm 0.1$ for small craters, close to the geometric saturation slope of $-2$ (\cite{Ma12}). This approximate inverse square power-law has also been observed on other planetary bodies in the solar system, for example on Mars (see, e.g., \cite{WeTa11} and references therein). Since size-frequency distributions of impact craters are a powerful tool in planetology to date surfaces and reconstruct planetary evolution (see, e.g., \cite{MiNe10}, and the review by \cite{HaNe01} for Mars), it is important to understand their origin. The first detailed statistical model of impact cratering was given by \cite{Ma64} in the context of lunar exploration. Assuming that craters are circular but may overlap, are Poisson-distributed over the lunar surface subject to a time-dependent production function with probability density $p(r,t)$ for a crater radius to lie in the interval $(r,r+\rmd r)$, he showed in a subsequent work that for any $p(r,t)\propto r^{-\gamma-1}, \ \gamma>2$, the cumulative frequency of craters larger than radius $r$ is indeed approximately proportional to $r^{-2}$ in a steady-state at late times (\cite{Ma66}). Therefore, when this state is reached, the distribution of craters is the same for a class of underlying production functions. Moreover, if a population of craters of radius $r$ reaches geometric saturation, then its frequency is exactly proportional to $r^{-2}$ because of circle packing on the surface (cf. \cite{Ga70}). It may also be recalled that power-law distributions are self-similar (for craters see, e.g., \cite{Ta90},  pp. 18f and 34ff), but this feature alone does not explain the fairly consistent power of $-2$ itself.

The universality of this property, then, seems to lend itself naturally to a geometrical explanation. In this article, we present a new geometrical approach to study crater distribution functions, which is inspired by applications of measures to cosmological models of inflation (e.g., \cite{GiTu08}). While any size-frequency distribution of craters is, in the absence of other resurfacing, the result of the production function which depends on properties of the impactors creating the craters, it is possible that the size-frequency distribution evolves to a steady-state at least partially independent of its creation history, as shown by Marcus' result. In such a case, one might expect \textit{a priori} a uniform distribution over the configuration space of craters. In order to investigate this point, we shall start in this paper with the assumption of a uniform distribution of craters over their configuration space and show that this implies in fact an inverse square power-law for the size-frequency distribution of small craters. Of course, by contraposition, this means that any deviation from this relationship implies a non-uniformity of the crater distribution over their configuration space, which in turn may have physically interesting implications for the underlying class of production functions.

The structure of this paper is as follows. In Sec. \ref{sec:desitter}, we consider the 3-dimensional configuration spaces of circles on the Euclidean plane and on the unit sphere, and it turns out that these can be given a de-Sitter geometry. By interpreting the configuration space of craters as that of circles and computing its canonical volume measure, we show in Sec. \ref{sec:craters} how an inverse square power-law for the cumulative size-frequency distribution of small craters emerges. This model also shows that the global crater distribution is no longer self-similar once crater size becomes comparable to the size of the planetary body. Initially, we shall assume that craters are sparse with negligible overlap in the sense that their area filling factor $F\ll 1$, corresponding to a small fraction of geometric saturation which has filling factor $\pi/(2\sqrt{3})\approx0.907$ (cf. \cite{Ga70}). This appears to be a reasonable assumption, as the observed distribution of craters is typically below 10\% of geometric saturation (e.g. for the Moon, see again \cite{Ga70} who compares it to a ''Mare Exemplum" created in laboratory experiments). The effect of crater overlap on the distribution function is then considered as well. The conclusions are in Sec. \ref{sec:concl}, where an interpretation of this result and further applications are suggested.

\section{De-Sitter circle geometry}
\label{sec:desitter}
We begin by showing that the geometry of the configuration spaces of circles on the Euclidean plane and on the unit sphere in Euclidean space can be regarded as de-Sitter spacetimes. The first case will be seen in Sec. \ref{sec:plane} as special case of the configuration spaces of $n$-spheres in $n+1$-dimensional Euclidean spaces. The discussion of this more general case will also help to establish the second case in Sec. \ref{sec:sphere}. Then, in both cases, one can easily derive the canonical volume measures for these configuration spaces. The connection with more general sphere geometry is outlined in the Appendix.
\subsection{On the plane}
\label{sec:plane}
Consider $n$-dimensional unoriented spheres $\mathbb{S}^n_r$ in the $n+1$-dimensional Euclidean space $\mathbb{E}^{n+1}$. Any such sphere is uniquely defined by its centre at $\mathbf{a}\in \mathbb{E}^{n+1}$ and its radius $r> 0$. Since these are independent of each other, the configuration space of such spheres $\mathcal{M}^{n+2}_E=\{\mathbb{S}^n_r\subset\mathbb{E}^{n+1}\}$ is $n+2$-dimensional. The special case $n=1$ corresponds, of course, to circles on the Euclidean plane. We will now see how the configuration space $\mathcal{M}^{n+2}_E$ can be interpreted as a $n+2$-dimensional de-Sitter spacetime, and this spacetime of constant curvature can be described in the usual way as a quadric hypersurface in $n+3$-dimensional Minkowski spacetime $\mathbb{E}^{1,n+2}$ with metric $\eta_{ab}=\mathrm{diag}(-1,1,...,1)$.

Firstly, one may regard the Euclidean space $\mathbb{E}^{n+1}$ as the intersection of the null cone through the origin in an $n+3$-dimensional Minkowski spacetime $\mathbb{E}^{1,n+2}$ with a null hypersurface, as follows. Given an arbitrary $\mathbf{x}\in \mathbb{E}^{n+1}$, define
\[
X^a=\left(\frac{\mathbf{x}^2+1}{2},\mathbf{x},\frac{\mathbf{x}^2-1}{2}\right)\in \mathbb{E}^{1,n+2},
\]
where the square of vectors in $\mathbb{E}^{n+1}$ is with respect to the corresponding Euclidean metric, and let
\[
N^a=(1,0,...,0,1) \in \mathbb{E}^{1,n+2}.
\]
Then each $\mathbf{x}$ is identified with a point which is both on the null cone through the origin,
\[
\eta_{ab}X^aX^b=0,
\]
and on the hypersurface
\[
\eta_{ab}X^aN^b=-1, 
\]
which is null since $\eta_{ab}N^aN^b=0$.

Secondly, consider a spacelike vector $Y^a=(Y^0,\mathbf{Y},Y^{n+2})\in\mathbb{E}^{1,n+2}$, so that $\eta_{ab}Y^aY^b>0$, which is perpendicular to $X^a$,
\begin{equation}
\eta_{ab}X^a Y^b=0.
\label{sphere1}
\end{equation}
Writing $Y\cdot N=\eta_{ab}Y^aN^b=-Y^0+Y^{n+2}$ for short, one can recast condition (\ref{sphere1}) as
\begin{equation}
\left(\mathbf{x}+\frac{\mathbf{Y}}{Y\cdot N}\right)^2=\frac{\eta_{ab}Y^aY^b}{(Y\cdot N)^2}. 
\label{sphere2}
\end{equation}
So for any $k\neq0$, a given $k Y^a$ uniquely defines an $n$-sphere in $\mathbb{E}^{n+1}$ with centre and radius
\[
\mathbf{a}=-\frac{\mathbf{Y}}{Y\cdot N}, \qquad r=\sqrt{\frac{\eta_{ab}Y^aY^b}{(Y\cdot N)^2}}, 
\]
respectively. Because of the freedom to rescale $Y^a$ we may set, without loss of generality,
\begin{equation}
 \eta_{ab}Y^aY^b=1,
\label{desitter1}
\end{equation}
and use (\ref{desitter1}) to express $Y^a$ in terms of the variables defining the corresponding sphere,
\begin{eqnarray}
Y^0 &=&-\frac{1}{2}\left(\frac{\mathbf{a}^2+1}{r}-r\right), \label{desitter2a}\\
\mathbf{Y} &=& -\frac{\mathbf{a}}{r}, \\
Y^{n+2}&=& -\frac{1}{2}\left(\frac{\mathbf{a}^2-1}{r}-r\right). \label{desitter2b}
\end{eqnarray}
Hence, this configuration space of spheres in Euclidean space is seen to be an $n+2$-dimensional 
de-Sitter spacetime $dS^{n+2}$,
\[
\mathcal{M}^{n+2}_E=\{Y^a\in\mathbb{E}^{1,n+2}: \eta_{ab}Y^aY^b=1\}=dS^{n+2},
\]
as required. The metric on this de-Sitter spacetime is induced from the Minkowski metric
of $\mathbb{E}^{1,n+2}$,
\begin{eqnarray}
\rmd s^2 &=& -(\rmd Y^0)^2+\rmd\mathbf{Y}^2+(\rmd Y^{n+2})^2 \nonumber\\
\phantom{} &=& \frac{1}{r^2}(-\rmd r^2+\rmd\mathbf{a}^2)\label{desitter3},
\end{eqnarray}
using (\ref{desitter2a}--\ref{desitter2b}). With $g^E_{ab}$ as the metric of the de-Sitter spacetime, one can rewrite (\ref{desitter3}) as
\[
 \rmd s^2=g^E_{ab}\rmd y^a \rmd y^b, \qquad y^a=(r,\mathbf{a})\in dS^{n+2},
\]
to obtain
\begin{eqnarray}
\rmd V^{n+2}_E&=&\sqrt{-\det{g^E_{ab}}}\rmd r\rmd a^1\ldots \rmd a^{n+1}\nonumber\\
\phantom{}&=&\frac{1}{r^{n+2}}\rmd r\rmd a^1\ldots \rmd a^{n+1}
\label{desitter4}
\end{eqnarray}
as the volume measure of the de-Sitter and hence the configuration space. Finally, as a special case of Eq. (\ref{desitter4}) one finds
\begin{equation}
\rmd V^3_E=\frac{1}{r^3}\rmd r\rmd a^1\rmd a^2, 
\label{onplane}
\end{equation}
the volume measure of the 3-dimensional configuration space $\mathcal{M}^{3}_E$ of circles on the Euclidean plane.
\subsection{On the sphere}
\label{sec:sphere}
In view of the application to the distribution of craters on a planetary surface, we shall proceed by extending the previous case to the configuration space of circles $\mathbb{S}^1_\rho$ on the 2-dimensional unit sphere $\mathbb{S}^2$ centered at the origin of $\mathbb{E}^3$. Each circle is fully characterized by its angular radius $\rho\in (0,\pi/2)$ and by its centre $\mathbf{n}$ on the unit 2-sphere in Euclidean 3-space, so that $\mathbf{n}^2=1$ and the two spherical coordinates of the centre can be computed in the usual way. The corresponding configuration space $\mathcal{M}^3_S$ is, as before, clearly 3-dimensional and can also be considered a de-Sitter spacetime.

\begin{figure}
\includegraphics[width=0.55\columnwidth]{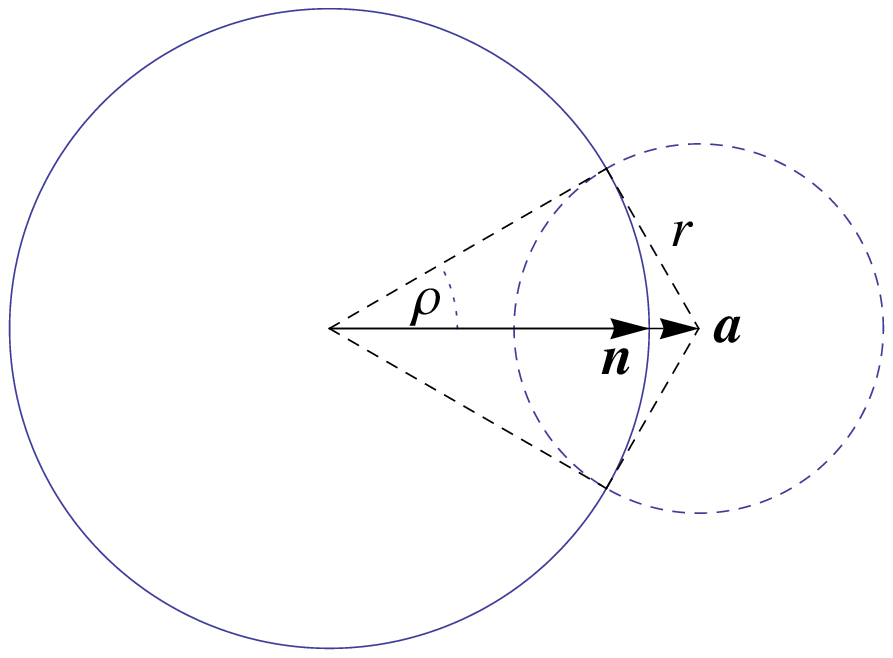}
\includegraphics[width=0.40\columnwidth]{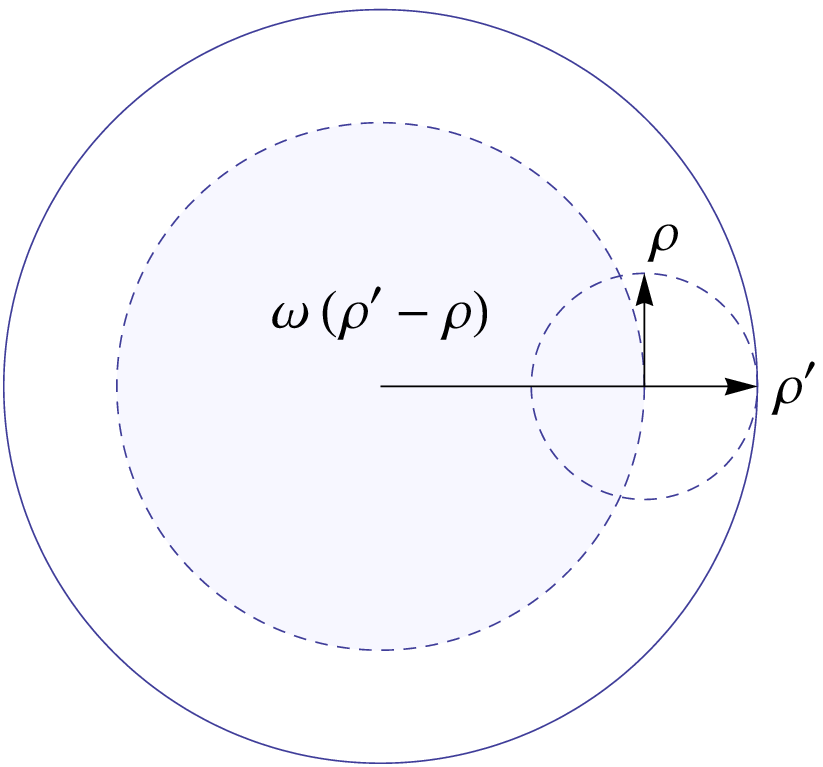}
\caption{\textbf{Left:} A circle $\mathbb{S}^1_\rho$ of angular radius $\rho$ at $\mathbf{n}$ on the unit 2-sphere $\mathbb{S}^2$ (shown as solid circle) in Euclidean 3-space is represented as the intersection of a 2-sphere $\mathbb{S}^2_r$ of radius $r=\tan \rho$ at $\mathbf{a}=\mathbf{n}\sec\rho$ (shown as dashed circle) with the unit 2-sphere. \textbf{Right:} The centre of a crater of angular radius $\rho$ wholly overlapped by a crater of angular radius $\rho'$ may lie within the shaded area of solid angle $\omega(\rho'-\rho)$.}
\label{fig:craters}
\end{figure}

To see this easily, one can utilize a trick to discuss this problem in terms of the previous one. Each circle on our unit 2-sphere $\mathbb{S}^2$ may be regarded as the intersection of another 2-sphere $\mathbb{S}^2_r$ of radius $r$ centred at $\mathbf{a}\in \mathbb{E}^3$ with $\mathbb{S}^2$. As noted in the previous section, the configuration space $\mathcal{M}^{4}_\mathbb{E}$ of 2-spheres in Euclidean 3-space is a 4-dimensional de-Sitter spacetime defined by the quadric (\ref{desitter1}) in 5-dimensional Minkowski spacetime $\mathbb{E}^{1,4}$ with coordinates (\ref{desitter2a}--\ref{desitter2b}). Now, for a given circle of angular radius $\rho$, there is a family of spheres $\mathbb{S}^2_r$ whose intersection with the unit 2-sphere produces this circle. Hence, their radius $r$ depends on the choice of $\mathbf{a}$ which, by symmetry, must be parallel to $\mathbf{n}$. As shown in Fig. \ref{fig:craters} (left panel), we can choose $\mathbf{a}$ such that $r=\tan \rho$ and $\mathbf{a}=\mathbf{n}\sec \rho$. Then from (\ref{desitter2a}--\ref{desitter2b}), one finds $Y^4=0$ and
\begin{eqnarray}
Y^0 &=&-\cot\rho, \label{desitter5a}\\
\mathbf{Y} &=& -\mathbf{n}\csc \rho. \label{desitter5b}
\end{eqnarray}
With this choice, we have coordinates $Z^a=(Y^0,\mathbf{Y})\in \mathbb{E}^{1,3}$ on the 4-dimensional Minkowski subspace $Y^4=0$ which determine uniquely the centre and angular radius of a circle on the unit 2-sphere and, since $\eta_{ab}Z^aZ^b=-(Y^0)^2+\mathbf{Y}^2=1$, define a 3-dimensional de-Sitter spacetime as the corresponding configuration space,
\[
\mathcal{M}^{3}_S=\{Z^a\in\mathbb{E}^{1,3}: \eta_{ab}Z^aZ^b=1\}=dS^{3},
\]
as promised. The metric on this de-Sitter spacetime is again induced by the ambient Minkowski space $\mathbb{E}^{1,3}$,
\[
\rmd s^2 = -(\rmd Y^0)^2+\rmd\mathbf{Y}^2=\csc^2\rho(-\rmd\rho^2+\rmd\mathbf{n}^2),
\]
from (\ref{desitter5a}--\ref{desitter5b}), subject to $\mathbf{n}^2=1$. One can write the line element on the unit sphere in terms of the usual spherical coordinates $(\theta,\phi)$,
\[
\rmd\mathbf{n}^2=\rmd\theta^2+\sin^2 \theta \rmd\phi^2,
\]
and thus express the line element of the configuration space as
\[
 \rmd s^2=g^S_{ab}\rmd z^a \rmd z^b, \qquad z^a=(\rho,\theta,\phi)\in dS^3,
\]
with metric
\begin{equation}
g^S_{ab}=\mathrm{diag}\left(-\csc^2\rho,\csc^2\rho,\csc^2\rho\sin^2\theta\right). 
\label{desitter6}
\end{equation}
Thinking of a sphere in $\mathbb{E}^3$ of radius $R_0$ with a circle of radius $R$ measured on the spherical surface so that its angular radius is $\rho=R/R_0$, then this metric (\ref{desitter6}) clearly reduces to the one in Eq. (\ref{desitter3}) in the limit of small circles, $\rho\ll 1$, as expected. The volume measure corresponding to metric (\ref{desitter6}) becomes
\begin{eqnarray}
 \rmd V^3_S&=&\sqrt{-\det g^S_{ab}}\rmd\rho \rmd\theta \rmd\phi \nonumber \\
 \phantom{}&=&\csc^3\rho \sin\theta \rmd\rho \rmd\theta \rmd\phi=\csc^3\rho\rmd\rho \rmd\omega, \label{onsphere}
\end{eqnarray}
for the 3-dimensional configuration space $\mathcal{M}^3_S$ of circles on the unit sphere, where $\rmd\omega=\sin\theta \rmd\theta \rmd\phi$ is the solid angle element on the unit sphere.

\section{Application to crater distributions}
\label{sec:craters}
Having determined the geometry of the configuration space $\mathcal{M}^3_E$ of circles on the Euclidean plane and $\mathcal{M}^3_S$ of circles on the unit sphere, we shall now show how this may be applied to the distribution of craters on planetary surfaces. The assumptions of our model shall be that, firstly, each crater can be represented by a circle and, secondly, the distribution of circles is uniform on their configuration space as defined in Sec. \ref{sec:desitter}. The latter assumption therefore serves as a null hypothesis about the structure of the configuration space of craters. However, the configuration spaces of circles allow arbitrary overlap whereas, naturally, this is not the case for craters erasing older and smaller ones within. In Sec. \ref{sec:sparse}, we shall consider the case that craters are sparse so that their overlap may be ignored. This condition is relaxed in Sec. \ref{sec:overlap}, where we attempt to indicate the effect of overlap on the distribution function of craters.   
\subsection{Sparse craters}
\label{sec:sparse}
\subsubsection{On the plane}
Suppose that an infinitesimal area $\rmd A$ in the Euclidean plane contains the centres of $\rmd n$ craters of radii in the interval $[r, r+\rmd r]$. Then the corresponding distribution function $f_E$ may be written $\rmd n=f_E \rmd r \rmd A$ so that, given our assumptions and the result from Sec. \ref{sec:plane}, $\rmd n \propto \rmd V^3_E$ and hence
\begin{equation}
f_E=\frac{C_E}{r^3} 
\label{f-e}
\end{equation}
from Eq. (\ref{onplane}), where $C_E$ is a constant. If this constant is sufficiently small, the assumption of sparse craters will hold. In order to determine this and estimate its value, let $n_E$ be the total number of craters with radius $r_{\min}\leq r\leq r_{\max}$ wholly within a disk of radius $R\geq r_{\max}$. Then the area filling factor of these craters is
\[
F_E=\int_{r_{\min}}^{r_{\max}} \pi r^2 f_E(r) \rmd r=\pi C_E \ln\frac{r_{\max}}{r_{\min}}
\]
from (\ref{f-e}), so that for sparse craters with $F_E \ll 1$ we must have
\begin{equation}
C_E\ll\left(\pi \ln \frac{r_{\max}}{r_{\min}}\right)^{-1}.
\label{sparse-e}
\end{equation}
As $R\rightarrow \infty$ and the condition (\ref{sparse-e}) continues to hold, $n_E/\pi R^2$ tends to the total frequency of craters with this radial range per unit area on the Euclidean plane,
\[
\nu_E=\int_{r_{\min}}^{r_{\max}} f_E \rmd r,
\]
whence, using again (\ref{f-e}), 
\begin{equation}
C_E=2\nu_E r_{\min}^2 \left(1-\frac{r^2_{\min}}{r^2_{\max}}\right)^{-1}.
\label{c-e}
\end{equation}
The cumulative frequency $\nu (r)$ of craters with radius greater than $r$ per unit area as a function of size can also be obtained from the distribution function (\ref{f-e}),
\begin{equation}
\nu(r)=\int_r^{r_{\max}}f_{E}(r')\rmd r'=\frac{C_E}{2}\left(\frac{1}{r^2}-\frac{1}{r^2_{\max}}\right).
\label{nu}
\end{equation}
Finally, note that the approximation $r_{\min}\ll r_{\max} \ll R$ applied to Eq.s (\ref{sparse-e}--\ref{nu}), which is reasonable for a crater count, gives $C_E \approx 2\nu_E r^2_{\min}\ll 1$ and an inverse square power-law for the cumulative frequency,
\begin{equation}
\nu(r)\approx \nu_E\frac{r_{\min}^2}{r^2}.
\label{nu-approx}
\end{equation}

\subsubsection{On the sphere}
\begin{figure}
\includegraphics[width=\columnwidth]{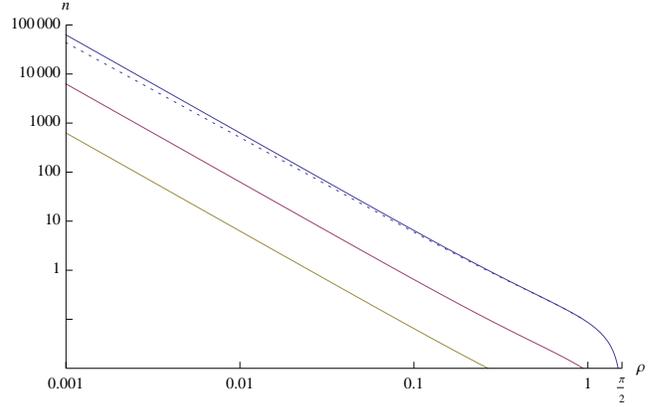}
\caption{Double logarithmic plot of the total cumulative number $n$ of craters larger than angular radius $\rho$ under the assumption of sparse craters (solid curves), for $\rho_{\min}=10^{-3}$ and three values of the constant in Eq. (\ref{n}): from top to bottom $C_S= 10^{-2}, 10^{-3},10^{-4}$, respectively. Note that the curves approach the inverse square power-law $n\propto \rho^{-2}$ for small $\rho$. The dotted curve shows the effect of crater overlap for the top curve with $\lambda=-0.1$ in Eq. (\ref{overlap4}).}
\label{fig:cumul}
\end{figure}

When the maximum radius of the craters is not negligible compared to the radius of the planetary body then, given our assumptions, the configuration space of circles on the unit sphere from Sec. \ref{sec:sphere} is applicable. Let $R_0$ be the radius of the planetary body and $R$ be the radius of a crater measured on the spherical surface so that its angular radius is $\rho=R/R_0\in (0,\pi/2)$ as before. Now suppose $\rmd n$ is the number of the centres of craters with angular radius in the interval $[\rho,\rho+\rmd\rho]$ within the solid angle element $\rmd \omega=\rmd A/R_0^2$, so $\rmd n=f_S\rmd \rho \rmd \omega$ with distribution function $f_S\propto\rmd V^3_S$ given by
\begin{equation}
f_S(\rho)=C_S\csc^3 \rho,
\label{f-s}
\end{equation}
from (\ref{onsphere}), where $C_S$ is a constant. A crater of angular radius $\rho$ subtends a solid angle 
\begin{equation}
\omega(\rho)=\int_0^{2\pi}\int_0^\rho \sin \theta \rmd \theta \rmd \phi=2\pi(1-\cos \rho)
\label{solid}
\end{equation}
on the planetary surface, so the area filling factor of craters with angular radius $\rho_{\min}\leq r\leq \pi/2$ is then
\begin{eqnarray*}
F_S&=&\int_{\rho_{\min}}^{\frac{\pi}{2}}\omega(\rho)f_S(\rho)\rmd \rho \\
\phantom{}&=&\pi C_S\left(\frac{\cos\rho_{\min}}{1+\cos\rho_{\min}}+\ln\cot\frac{\rho_{\min}}{2}\right).
\end{eqnarray*}
For the assumption of sparse craters to hold, we require that $F_S\ll 1$ as before. Now if $\rho_{\min}\ll 1$, that is, the smallest craters counted are much smaller than the radius of the planetary body, one must therefore require that
\begin{equation}
C_S \ll \left(\frac{\pi}{2}+\pi\ln\frac{2}{\rho_{\min}}\right)^{-1}.
\label{sparse-s}
\end{equation}
The total number of craters with radius greater than $\rho_{\min}$ is
\begin{eqnarray}
 n_S&=&\int_0^{4\pi}\int_{\rho_{\min}}^{\frac{\pi}{2}}f_S(\rho)\rmd\rho\rmd \omega \nonumber\\
\phantom{} &=& 2\pi C_S \left(\cot \rho_{\min} \csc \rho_{\min}+ \ln \cot \frac{\rho_{\min}}{2}\right).\label{n-tot}
\end{eqnarray}
Again if $\rho_{\min}\ll 1$, then we can approximate (\ref{n-tot}) by the leading monomial to find
\begin{equation}
C_S \approx \frac{n_S \rho_{\min}^2}{2\pi}.
\label{c-s}
\end{equation}
The total cumulative number of craters with radius greater than $\rho$ is also obtained from (\ref{f-s}),
\begin{eqnarray}
n(\rho)&=&4\pi \int_\rho^{\frac{\pi}{2}} f_S(\rho') \rmd \rho'  \nonumber \\
&=&2\pi C_S\left(\cot \rho \csc \rho+ \ln \cot \frac{\rho}{2}\right).
\label{n}
\end{eqnarray}
This function is shown in Fig. \ref{fig:cumul} for three values of $C_S$, for which condition (\ref{sparse-s}) holds in the range shown. Also, if $\rho_{\min}\ll 1$, one can use the approximation (\ref{c-s}) for the constant in Eq. (\ref{n}) to find 
\begin{equation}
n(\rho)\approx n_S \rho_{\min}^2\left(\cot \rho \csc \rho+ \ln \cot \frac{\rho}{2}\right).
\label{n-approx}
\end{equation}
This expression clearly recovers Eq. (\ref{nu-approx}) in the limit of $\rho\ll 1$ where $4\pi R_0^2\nu=n$ and $\rho R_0=R\approx r$, as expected. Likewise, one can easily convert (\ref{n}) to the cumulative crater frequency per unit area of the planetary body as a function of crater size. Note also that the global cumulative number (\ref{n}) is given by trigonometric functions and not a power-law. Hence, the crater distribution is no longer self-similar, as is the case for small craters or for craters on the Euclidean plane in the limit $r_{\max}\rightarrow \infty$ of Eq. (\ref{nu}).
\subsection{Overlapping craters}
\label{sec:overlap}
The spherical model discussed in Sec. \ref{sec:sphere} is, of course, the generalization of the planar model of Sec. \ref{sec:plane} applicable to sparse craters on a planetary surface. However, a more realistic model for the distribution of craters should take into account the possibility of overlap. In this section, we shall consider how the more general spherical model can thus be extended. For simplicity we shall assume that, firstly, there is a finite period of crater formation and, secondly, the crater distribution on the planetary surface is always isotropic. Thirdly, we take (\ref{f-s}) to be the intrinsic distribution  of craters produced, which may differ from the distribution observed on the planetary surface due to overlap. Finally, we assume that the centre and radius of a crater can be identified as long as a sector of its perimeter survives, that is, only craters that are wholly overlapped by a later and larger crater are considered to be obliterated.

Hence, one may take the initial distribution to be $f_S(\rho)$ and the final observed distribution $f_O(\rho)$ to be a function of $\rho$ only. Craters of radius $\rho$ will have been wholly overlapped by larger craters of radius $\rho'>\rho$ if their centres are within a solid angle of
\begin{equation}
\omega(\rho'-\rho)=2\pi(1-\cos(\rho'-\rho))
\label{solid2}
\end{equation}
from Eq. (\ref{solid}), concentric with a larger crater of radius $\rho'$ in the observed distribution, of which there are $4\pi f_O(\rho')\rmd \rho'$. This is illustrated in Fig. \ref{fig:craters} (right panel). In addition, there remain $4\pi f_O(\rho)\rmd \rho$ craters of radius $\rho$ in the distribution observed on the planetary surface. By assumption, we take the distribution of craters of radius $\rho$ that are produced to be proportional to $f_S(\rho)$. Then this applies to the overlapped craters as well as to the sum of the observed craters plus the overlapped ones, of course with different constants of proportionality. We therefore expect to have
\begin{equation}
f_O(\rho)+\int_\rho^{\frac{\pi}{2}}f_O(\rho')f_S(\rho)\omega(\rho'-\rho)\rmd \rho' \propto f_S(\rho).
\label{overlap0}
\end{equation}
It may be useful to express the distribution function as
\[
f_O(\rho)=C_O f(\rho)
\]
where $C_O$ is the scaling factor proportional to the total number of observed craters $n_O$ greater than $\rho_{\min}$, which reduces to $C_S$ in the limit of sparse craters,
\begin{equation}
n_O=4\pi C_O \int_{\rho_{\min}}^{\frac{\pi}{2}}f(\rho)\rmd \rho.
\label{n-obs}
\end{equation}
Hence, using (\ref{f-s}) and (\ref{solid2}), one can rewrite (\ref{overlap0}) as
\begin{equation}
f(\rho)=g(\rho)+\lambda \int_\rho^{\frac{\pi}{2}} f(\rho') K(\rho,\rho') \rmd \rho',
\label{overlap1}
\end{equation}
where
\[
g(\rho)=\csc^3\rho,
\]
and
\[
K(\rho,\rho')=\csc^3\rho (1-\cos(\rho'-\rho)),
\]
and $\lambda<0$ is a (by definition negative) constant. The problem of overlapping craters, at least in this simplified form, can therefore be expressed as the integral equation (\ref{overlap1}), which is the standard form of a linear, inhomogeneous Volterra equation of the second kind. Note that its kernel is of separable form since it may be written
\[
K(\rho,\rho')=\sum_i k_i(\rho)k'_i(\rho'),
\]
where the $k_i$ and $k'_i$ are functions only of $\rho$ and $\rho'$, respectively. Alternatively, one can turn the integral equation (\ref{overlap1}) into a differential equation. Letting
\[
\tilde{f}(\rho)=\frac{f(\rho)}{g(\rho)}, 
\]
we get
\begin{equation}
\lambda g\tilde{f} +\frac{\rmd\tilde{f}}{\rmd \rho}+\frac{\rmd^3 \tilde{f}}{\rmd \rho^3}=0,
\label{overlap2}
\end{equation}
a linear, homogeneous ordinary differential equation of third order with non-constant coefficients. Furthermore, one can define
\[
\hat{f}(\rho)=\frac{\rmd \ln \tilde{f}}{\rmd \rho}
\]
to reduce the order of (\ref{overlap2}) and convert it to
\begin{equation}
\lambda g+\hat{f}+\hat{f}^3+3\hat{f}\frac{\rmd \hat{f}}{\rmd \rho}+\frac{\rmd^2 \hat{f}}{\rmd \rho^2}=0,
\label{overlap3}
\end{equation}
a non-linear, inhomogeneous ordinary differential equation of second order with constant coefficients. While a detailed discussion of the properties of Eq.s (\ref{overlap1})--(\ref{overlap3}) and the numerical comparison with observed crater distributions is beyond the scope of the present article, we shall use the Volterra equation (\ref{overlap1}) to illustrate the effect of \textit{little overlap} for craters with small lower limit of the angular radius, $\rho_{\min} \ll 1$. By this we mean a small deviation of $f_O$ from the sparse crater distribution so that the constant in the Volterra equation is small, $|\lambda|\ll 1$. Then an approximate solution of Eq. (\ref{overlap1}) for $f$ can be expressed naturally as the leading terms of a Neumann series expansion of the full solution,
\[
 f(\rho)=\sum_{j=0}^\infty \lambda^j u_j(\rho),
\]
whence the first approximation will be given by
\begin{eqnarray}
f(\rho)&\approx& u_0+\lambda u_1 \nonumber \\
\phantom{}&=&\csc^3 \rho+\lambda \int_{\rho}^{\frac{\pi}{2}}\csc^3\rho'\csc^3\rho(1-\cos(\rho'-\rho))\rmd \rho' \nonumber \\
\phantom{}&=&\csc^3\rho \left(1+\frac{\lambda}{2}\left(-\cos\rho+\ln\cot\frac{\rho}{2}\right)\right).
\label{overlap4}
\end{eqnarray}
Integrating (\ref{overlap4}) in Eq. (\ref{n-obs}), the leading monomial in $\rho_{\min}$ yields the approximation
\[
 n_O\approx\frac{2\pi C_O}{\rho_{\min}^2}\left(1-\frac{\lambda}{4}\left(3-2\ln\frac{2}{\rho_{\min}}\right)\right).
\]
Hence, the constant $\lambda$ indicates the importance of crater overlap. In the current formulation, it is a free parameter that will depend on the total number of craters produced and on the number $n_O$ of craters that remain observable.

\section{Conclusion}
\label{sec:concl}
Starting from the assumption that impact craters are uniformly distributed over their de-Sitter configuration space, it is shown by Eq.s (\ref{nu-approx}) and (\ref{n-approx}) that this can indeed explain the approximate inverse square power-law for the cumulative size-frequency distribution of small craters, which is widely observed as noted in the Introduction. Moreover, while Eq. (\ref{n}) shows that the global crater distribution is not self-similar, Fig. \ref{fig:cumul} illustrates that the deviation from the inverse square power-law due to the finite size of the planetary body is in fact mostly negligible, so that the crater distribution is effectively self-similar. Also, since smaller craters are relatively more affected by overlap, we expect that they are systematically undercounted relative to the distribution of sparse craters with negligible overlap. This is the case, as indicated by Eq. (\ref{overlap4}) and the dotted curve in Fig. \ref{fig:cumul} for small $\rho$.

Hence, the approach promoted in this article suggests that the de-Sitter configuration space may be a useful prior for the study of cratering histories. Deviations from the inverse square power-law correspond to non-uniform distributions of small craters on this configuration space. We suggest that such deviations from uniformity will indicate physically interesting processes like resurfacing due to igneous activity and erosion, as well as properties of the impactors before a steady-state of the crater population, as in the formalism of Marcus mentioned in the Introduction, is reached. A first step in this direction would be to classify production functions and resurfacing processes in terms of the deviation of the resulting crater configuration space from this de-Sitter geometry, which will depend on time. Hence, comparing these templates with observed maps of crater configuration spaces may yield new insights into cratering histories.

Finally, this article notes a perhaps surprising formal connection between the geometry of a cosmological spacetime and a problem in planetology. Recall that the 3-dimensional configuration space of circles, and hence of craters as discussed, was interpreted as the  3-dimensional de-Sitter spacetime with the metric given by (\ref{desitter6}) and hence the line element,
\[
\rmd s^2=- \csc^2 \rho \rmd \rho^2+\csc^2 \rho (\rmd \theta^2+ \sin^2\theta \rmd \phi^2).
\]
Now in order to connect this with the de-Sitter spacetime familiar from cosmology, one can use the coordinate transformation $\csc \rho=\cosh t$ so that
\[
\rmd s^2=-\rmd t^2+ \cosh^2 t (\rmd \theta^2+ \sin^2\theta \rmd \phi^2),
\]
where $t$ takes its usual interpretation as coordinate time. Extending our argument from two to three spatial dimensions, one obtains a measure on the space of spheres in 3-dimensional Euclidean space from Eq. (\ref{desitter4}) for $n=2$. This may be useful to study the statistics of bubbles in the universe, for instance supernova remnants within the interstellar medium or voids in the large-scale structure. Also, eternal inflation treats of bubble universes which are expected to collide (see, e.g., \cite{GaGuVi07}), so finding a version of the overlap formula (\ref{overlap1}) applicable to this problem may further elucidate the collision process of bubble universes.      

\section*{Acknowledgments}
MCW gratefully acknowledges support by World Premier International Research Center Initiative (WPI Initiative), MEXT, Japan, and thanks Dr Steven Gratton for discussions at the 2012 Kavli Astrophysics Symposium, Chicheley Hall, UK.

\section*{Appendix}
Here we give a brief discussion of the geometry underlying the construction in Sec. \ref{sec:desitter}. 
Circle geometry, beginning with the classical problem of Apollonius to find the circles touching three given ones, is a special case of 
the sphere geometries developed by Lie (1872) and Laguerre (1881). Following partially the treatment by \cite{Be12}, consider the Euclidean 
space $\mathbb{E}^{n+1}$ with its standard Euclidean metric. Then a Laguerre cycle is either a point or an oriented $n$-sphere of 
radius $r\neq 0$ centered at $\mathbf{a}\in \mathbb{E}^{n+1}$, with orientation expressed by the sign of $r$. Laguerre cycles can thus be regarded 
as elements of a real vector space $V=\mathbb{R}\oplus\mathbb{E}^{n+1}$. Assigning coordinates $c=(r,\mathbf{a})$ to a Laguerre cycle is called cyclographic projection, and addition of Laguerre cycle coordinates is defined as for vectors. The zero Laguerre cycle is $0=(0,\mathbf{0})$. Now given two Laguerre cycles $c_1=(r_1,\mathbf{a}_1),\ c_2=(r_2,\mathbf{a}_2)$, the real number
\begin{equation}
P(c_1,c_2)=-(r_2-r_1)^2+(\mathbf{a}_2-\mathbf{a}_1)^2, 
\label{power}
\end{equation}
where the product of vectors in $\mathbb{E}^{n+1}$ is understood to be the inner product with respect to the Euclidean metric, is called power of $c_1$ and $c_2$. It turns out that $c_1$ and $c_2$ touch each other respecting orientation if, and only if, (\cite{Be12}, proposition 3.43)
\[
P(c_1,c_2)=0,
\]
and this is exactly the reflexive and symmetric contact relation studied in sphere geometry. Moreover, $V$ can be equipped with a product defined as
\[
c_1 c_2=-r_1r_2+ \mathbf{a}_1 \mathbf{a}_2.
\]
Letting $\rmd c=(\rmd r, \rmd\mathbf{a})$, we therefore find that
\begin{equation}
\rmd c^2=-\rmd r^2+\rmd \mathbf{a}^2=P(c,c+\rmd c)
\label{lagmetric}
\end{equation}
measures the infinitesimal deviation of Laguerre cycles with respect to the contact relation, and this is seen to define a Minkowski metric on $V$. Note that (\ref{lagmetric}) may be regarded as a line element on $V$ which is invariant under transformations of Laguerre cycles preserving power (\ref{power}). Hence, there is a close connection between the Minkowski spacetime of special relativity and the geometry of Laguerre cycles, which appears to have been pointed out first by \cite{Ti12}. In addition to the set of Laguerre cycles $\Gamma$, one can introduce the set of oriented Euclidean hyperplanes $\Sigma$ in $\mathbb{E}^{n+1}$ and the object infinity $\infty$, and define corresponding contact relations. Then $\Delta=\Gamma \cup \Sigma \cup \{\infty\}$ is called the set of Lie cycles, and bijections $\Lambda:\Delta\rightarrow\Delta$ that preserve contact relations form the group of Lie transformations of $\mathbb{E}^{n+1}$. It turns out (\cite{Be12}, proposition 3.56) that to every Lie cycle a homogeneous Lie cycle coordinate $L=[L^0,\mathbf{L}, L^{n+2},L^{n+3}]$, where $[L^0,\mathbf{L}, L^{n+2},L^{n+3}]=[kL^0,k\mathbf{L}, kL^{n+2},kL^{n+3}]$ for any $k\neq 0$, can be assigned bijectively, which satisfies the Lie quadric
\begin{equation}
-(L^0)^2+\mathbf{L}^2+(L^{n+2})^2-(L^{n+3})^2=0.
\label{lie}
\end{equation}
For a Laguerre cycle $c=(r,\mathbf{a})$, the Lie cycle coordinate is (c.f. \cite{Be12}, p. 154, applying a sign change and reordering)
\[
L(c)=\left[\frac{P(0,c)+1}{2},\mathbf{a}, \frac{P(0,c)-1}{2},-r\right].
\]
By homogeneity of the Lie cycle coordinates, we can set $L^{n+3}=1$ to obtain
\begin{eqnarray*}
L(c)&=&\left[-\frac{1}{2}\left(\frac{\mathbf{a}^2+1}{r}-r\right),-\frac{\mathbf{a}}{r}, -\frac{1}{2}\left(\frac{\mathbf{a}^2-1}{r}-r\right),1\right] \\
\phantom{}&=&\left[Y^0,\mathbf{Y},Y^{n+2},1\right],
\end{eqnarray*}
using (\ref{power}) and the coordinates  (\ref{desitter2a})-(\ref{desitter2b}) defined in Sec. \ref{sec:desitter}. Then these are seen to satisfy
\begin{equation}
-(Y^0)^2+\mathbf{Y}^2+(Y^{n+2})^2=1
\label{quadric}
\end{equation}
because of the Lie quadric (\ref{lie}). Suppose now that we consider the line element
\begin{eqnarray}
\rmd s^2&=&-(\rmd Y^0)^2+\rmd \mathbf{Y}^2+(\rmd Y^{n+2})^2 \label{minkowski} \\
\phantom{}&=& \frac{1}{r^2}(-\rmd r^2+\rmd \mathbf{a}^2)=\frac{\rmd c^2}{r^2} \nonumber
\end{eqnarray}
using (\ref{lagmetric}). Clearly, any transformation $(Y^0,\mathbf{Y},Y^{n+2})\mapsto (Y^0+\rmd Y^0,\mathbf{Y}+\rmd \mathbf{Y},Y^{n+2}+\rmd Y^{n+2})$ that preserves our choice (\ref{quadric}) leaves this line element invariant, and in fact null $\rmd s^2=0$ so that $\rmd c^2=0$. Hence, such transformations preserve the contact relations of the corresponding Laguerre cycles, as described above. Under this assumption, then, $(Y^0,\mathbf{Y},Y^{n+2})$ may be regarded as a point on the de-Sitter configuration space of Sec. \ref{sec:desitter} defined by the quadric hypersurface (\ref{quadric}) in the Minkowski spacetime $\mathbb{E}^{1,n+2}$ with line element (\ref{minkowski}). A connection between Laguerre sphere geometry and de-Sitter spacetime seems to emerge first in \cite{Gr34}, who uses the opposite metric signature.

\label{lastpage}
\end{document}